\documentclass[superscriptaddress, aps, prx, longbibliography]{revtex4-2}

\usepackage{graphicx}
\usepackage{physics}
\usepackage{amsmath}
\usepackage{array}
\usepackage{siunitx}
\usepackage[section]{placeins}
\usepackage{hyperref}
\usepackage{soul}
\usepackage{ulem}
\usepackage{cancel}

\usepackage{natbib}

\begin{document}

\title{Towards all-optical spin manipulation in single molecules: a refined region for locating a dark resonance}
\date{June 2025}

\author{Robert Smit}\email{rsmit@physics.leidenuniv.nl}
\affiliation{Huygens-Kamerlingh Onnes Laboratory, LION, Postbus 9504, 2300 RA Leiden, The Netherlands}
\author{Boleslaw Kozankiewicz}
\affiliation{Institute of Physics, Polish Academy of Sciences Al. Lotnikow 32/46, 02-668 Warsaw, Poland}
\author{Michel Orrit}
\affiliation{Huygens-Kamerlingh Onnes Laboratory, LION, Postbus 9504, 2300 RA Leiden, The Netherlands}

\begin{abstract}
The on-demand manipulation of triplet states in closed-shell single molecules is still out of reach due to a lack of information about the energy of those triplet states. Yet, the access to triplet states would open up a route towards an all-optical single-molecule photonic switch/transistor and, moreover, would provide a way of performing coherent spin operations from the spin-less ground state. In this work, we take an important step towards those aims by measuring the triplet energy from the weak phosphorescence signal of perdeuterated perylene, embedded as a guest molecule in a dibenzothiophene host matrix, which well preserves the coherence properties of the perylene guest. We find that perylene's phosphorescence can be enhanced in this host matrix, when acting as an intermediary for the generation of triplet excitons. The triplet energy that we find can be used to significantly narrow down the search for the ultra-weak spin-forbidden transitions from the ground singlet to the triplet states of a single molecule. 
\end{abstract}

\maketitle

\section{Introduction}

Spin-carrying systems provided by defects, such as nitrogen-vacancy (NV) centers in diamond\cite{Gruber1997Scanning}, lanthanide ions in garnets\cite{Zhong2015Optically} and the more recent discovered defects in hexagonal boron nitride\cite{Gottscholl2021Room}, have emerged as powerful platforms for quantum applications. As these systems possess ground electronic states with nonzero spin multiplicity, direct access to spin levels is allowed. The microwave-induced spin transitions in such systems can be detected optically through changes in the defect's fluorescence signal, knows as optically-detected magnetic resonance (ODMR) spectroscopy. Beyond microwave-driven transitions, fully optical schemes — demonstrated in both NV centers\cite{Santori2006Coherent} and atomic systems\cite{Nagourney1986Shelved} — enable spin-state manipulation without the need for microwaves. Spin-photon interfaces, relying on light-matter coupling, could then be used to enhance control over the spin state\cite{gurlek_small_2025}. \\
\\
The aforementioned spin systems have so far enabled a wide range of applications, including nanoscale magnetometry\cite{wolf_subpicotesla_2015}, thermometry\cite{ivady_pressure_2014}, demonstrations of quantum entanglement\cite{bernien_heralded_2013}, the development of quantum memories\cite{bradley_robust_2022} and of optomechanics\cite{delord_spin-cooling_2020}. While for quantum memories, electron spins can enable efficient coherent transfer of quantum information to a nuclear spin network through hyperfine interactions, the inability to turn off the electron spin after a write operation leaves a channel for decoherence for that memory\cite{gurlek_small_2025}. In contrast, systems without spin in their electronic ground state — such as closed-shell aromatic molecules — naturally avoid this issue. For this reason, molecular crystals have also been proposed as a potential solution in levitated optomechanics, namely in view of generating massive quantum states from a network of nuclear spins with long coherence times\cite{Steiner2024Pentacene}.\\
\\
Typical aromatic molecules feature a singlet ground state, a singlet excited state, and an intermediate triplet excited state (see Figure 1a) that can only be accessed by intersystem crossing from the singlet excited state. The triplet state comprises three spin sublevels ($T_x$, $T_y$, $T_z$), with their relative energies in zero magnetic field defined by zero-field splitting parameters $D$ and $E$\cite{miyokawa_zero-field_2024}. Following the discovery that particular single molecules embedded in host crystals can be detected by their fluorescence, magnetic resonance via ODMR was soon demonstrated for a pentacene molecule embedded in a \textit{p}-terphenyl matrix\cite{Wrachtrup1993Optical, Kohler1993Magnetic}. Subsequent works have resolved hyperfine couplings to nearby nuclear spins, such as ${}^{13}$C\cite{Kohler1995Single} and hydrogen\cite{Wrachtrup1997Magnetic}. While microwave-driven spin control is routine for single molecules, an all-optical counterpart — akin to those performed in NV centers and atomic systems — has yet to be achieved. With optical spin control, single-molecule fluorescence switches or transistors\cite{Orrit2009Photons} could be realized (see working principle in Figure 1b). Moreover, in view of quantum memories, optical spin control would provide a direct path towards storing quantum information, without having to rely on stochastic intersystem crossing before such an operation can be applied. \\

\begin{figure}
    \centering
    \includegraphics[scale=0.5]{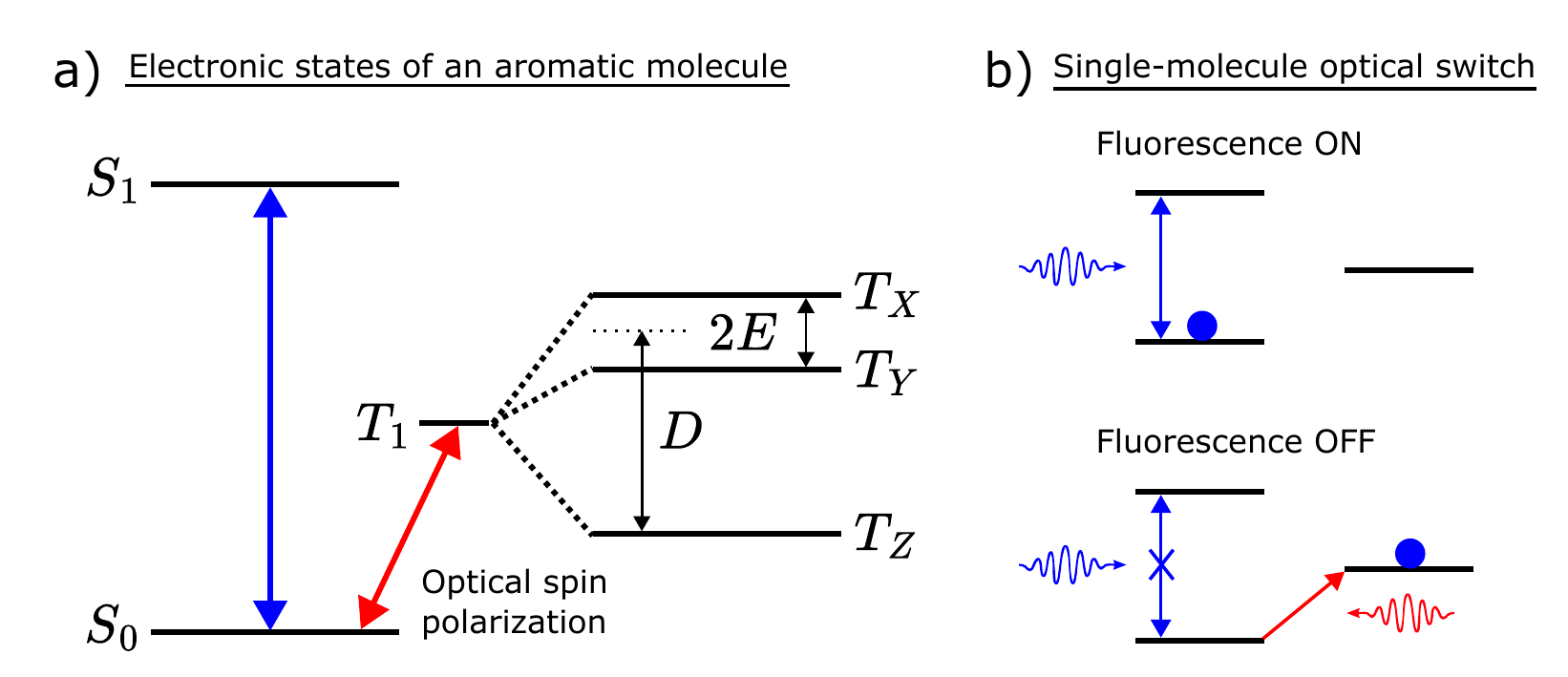}
    \caption{\textbf{The single-molecule optical switch}. The energy level diagram of a typical closed-shell aromatic molecule has a simplified structure as shown in panel (a). Panel (b) shows that driving the transition from the ground state $S_0$ to the metastable $T_1$ state (red arrows) allows switching off the fluorescence cycles between $S_0$ and $S_1$ (blue arrows), explaining the switch/transistor action.}
    \label{Figure1}
\end{figure}

Recent advances continue to stimulate research in the field of spin manipulations in molecules: novel experiments for example include electron spin resonance of pentacene detected via atomic force microscopy\cite{Sellies2023Single}, room-temperature ODMR of pentacene in \textit{p}-terphenyl\cite{Mena2024Room}, spin manipulation in optically-active organic radicals\cite{Gorgon2023Reversible} and experiments on new chemical species that have ODMR contrasts that well exceed those of NV centers\cite{mann_chemically_2025}. In this work, we take the first steps towards all-optical spin manipulation with perylene in a dibenzothiophene host crystal, by detecting the energy of its triplet from the emitted phosphorescence.  

\section{Methods}

\subsection{Preparation of perylene-d12 molecules in a dibenzothiophene crystal}

Perylene-d$_{12}$ (Sigma-Aldrich, 98\%) was dissolved in toluene (Acros Organics, 99.85\%) to prepare millimolar concentration solutions. Dibenzothiophene (Sigma-Aldrich), previously purified by zone refining, was melted using a hot-air gun. The perylene solution was then pipetted into the molten dibenzothiophene. After a brief period of additional heating to fully evaporate the toluene, the mixture was rapidly cooled, causing it to solidify. A small amount of the resulting solid mixture was pressed between two glass slides and reheated. The thin liquid film formed between the slides was then rapidly frozen in liquid nitrogen to ensure uniform dispersion of perylene molecules within the dibenzothiophene crystals. 

\subsection{Measurement setup }

The samples are measured in a liquid-helium flow cryostat (Janis, SVT-200-5) at a regulated temperature through control of the flow of liquid helium and pressure. The excitation is performed with a 100 W electrical power mercury lamp, placed in a housing (Oriel 60028) that boosts the light collection efficiency from the lamp and collimates the output beam. The ultraviolet part of the mercury emission spectrum (see Supplementary Figure S1) is transmitted by a filter (FSR-U340) and focused onto a feedback-controlled optical chopper wheel (Bentham 418F). The intensity-modulated light is then focused on the sample. The emitted light from the sample is subsequently collected from another window of the cryostat and directed to the opposite side of the chopper wheel. This is to ensure that the time windows of excitation and emission are not overlapped, but delayed. The duration of the delay between the end of the excitation window and the start of the acquisition window is controlled by the chopper frequency. The emitted light is then focused onto the slit of the spectrometer (Horiba IHR-320) with an attached Symphony II CCD detector or to a single-photon counting module (Excelitas, SPCM-AQRH-16). For the single-molecule experiments we made use of a home-built confocal setup with a Sacher diode laser (Littman/Metcalf) as a tunable excitation source, allowing us to work in the wavelength range of 452-457 nm at a spectral linewidth of 100 kHz. 

\section{Results and discussion}

Perylene guest molecules embedded in a dibenzothiophene host matrix constitute a host/guest system that exhibits narrow, highly stable spectral resonances at cryogenic temperatures, as demonstrated in our previous investigations of this material system\cite{Smit2022Reverse}. The presence of such narrow and stable resonances — effectively free from spectral diffusion — enables the isolation of individual molecular resonances from within an inhomogeneously broadened ensemble (see the zoom-in in Figure 2). This capability allows for prolonged measurements on single molecules, which is essential for obtaining stable fluorescence signals while searching for dark transitions to the triplet state. Such transitions would manifest as changes in the fluorescence rate, analogous to the optical detection of microwave-induced transitions between triplet spin sublevels.\\

\begin{figure}
    \centering
    \includegraphics[scale=0.35]{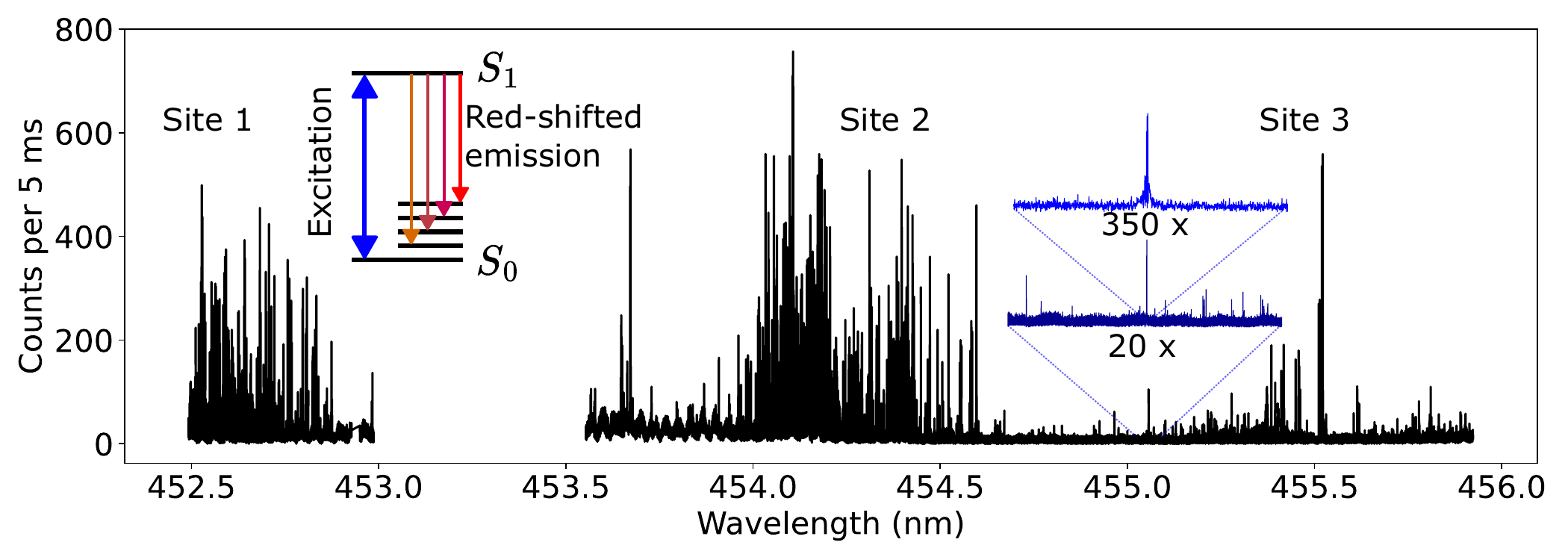}
    \caption{\textbf{Selecting single resonances from an inhomogeneous distribution}. The resonances from single perylene molecules in a dibenzothiophene crystal are scattered over a wide range of wavelengths and can be observed by collecting the red-shifted emission once the excitation laser hits resonance with a zero-phonon line. There are likely three distributions within the scanned range, where the part between 453 nm and 453.5 nm was skipped. The resolution of the scan is 6 MHz at an integration time of 5 ms per data point, making the total scan duration a minimum of 43 minutes - excluding times required to move the motor of the laser grating.}
    \label{Figure2}
\end{figure}

The inhomogeneous distribution of perylene molecules, as depicted in Figure \ref{Figure2}, indicates the presence of at least three distinct sub-distributions among the $>$527 detected molecules. These clusters, commonly referred to as spectroscopic sites, likely arise from variations in the local environment as molecules are incorporated into the host matrix, while other sources of structural inhomogeneity, such as isotope effects, contribute only marginally to the observed energy shifts\cite{Kohler1995Single}. A detailed examination of a single resonance, magnified by a factor of 350, reveals a characteristic linewidth of less than approximately $\sim$10$^{-5}$ nm ($\sim$100 MHz), approaching the natural, lifetime-limited linewidth of 45 MHz for this system\cite{Smit2022Reverse}. This linewidth is over five orders of magnitude narrower than the full scan range. In contrast, if one would search for a dark optical resonance to the triplet state, it must be taken into account that typical spin transitions, such as revealed by ODMR, are even a factor 100 $\times$ narrower\cite{Kohler1995Single} and thus making such a scan at a comparable resolution would probably take several days. During that same time, the molecule must stay at resonance with the probe laser to keep a steady fluorescence signal, as a reference for when the other laser hits resonance with a dark transition.\\
\\
The search for an ultra-narrow transition to the triplet state is therefore a considerable challenge, akin to locating a needle in a haystack. This search can be substantially refined if the approximate triplet-state energy for the molecular ensemble is known. While several methods exist for detecting triplet-state signals, direct absorption techniques are rendered impractical due to the extremely weak oscillator strengths associated with singlet–triplet transitions ($\sim$10$^{-10}$). Additionally, the generally low intersystem crossing rates and poor phosphorescence quantum yields characteristic of typical fluorophores employed in single-molecule spectroscopy (a comprehensive list is provided by Adhikari \textit{et al.}\cite{adhikari_future_2024}) further complicate direct detection.\\
\\
Perylene is a particularly attractive choice in this context, as it is among the bluest known single emitters, and according to the energy gap law, is predicted to possess a higher phosphorescence quantum yield. Notably, it remains the only molecule for which phosphorescence has been observed in cryogenic single-molecule spectroscopy experiments, namely when embedded within an anthracene host matrix\cite{Walla1998Perylene}. However, unlike perylene in dibenzothiophene, single-molecule detection was not achievable in that system, later attributed to intermolecular intersystem crossing effects\cite{Nicolet2006Intermolecular}. While the triplet energy of perylene has been determined in anthracene (12,844 cm$^{-1}$) and in pure crystalline form (12,372 cm$^{-1}$), these values can only serve as approximate references for dibenzothiophene, as solvent-induced shifts can significantly alter the electronic energy levels of guest molecules across different host matrices.\\
\\
The separation of the ultra-weak phosphorescence signal from the much stronger fluorescence background can be achieved through time-gated detection. Given that the fluorescence lifetime of perylene is on the order of nanoseconds, while the triplet-state lifetimes extend into the millisecond regime\cite{Smit2022Reverse}, time-resolved measurements allow effective discrimination between these emission processes. However, the yield of excited triplets remains relatively low, with only approximately 1 in 10$^{6}$ excited singlet states undergoing intersystem crossing to the triplet manifold\cite{Verhart2016Intersystem, Smit2022Reverse}. According to the energy gap law\cite{Englman1970energy} and considering a typical radiative lifetime on the order of 30 seconds\cite{Kellogg1964Radiationless,Siebrand1967Radiationless}, the phosphorescence quantum yield is expected to be approximately 10$^{-4}$. To enhance this yield, we employed perdeuterated perylene in place of the protonated analogue, as deuteration is known to modestly reduce the internal conversion rate\cite{Siebrand1967Isotope}.\\
\\
Nonetheless, not all excitation photons are absorbed, and only a minute fraction of the resulting phosphorescence photons are detected in our setup. Accounting for all contributing factors, we estimate an upper limit for the detection probability of phosphorescence photons per singlet excitation event to be on the order of a staggeringly low 10$^{-13}$. Increasing the excitation power is not a viable solution, as higher intensities promote reverse intersystem crossing, leading to an increased non-radiative decay from the triplet states in this host–guest system\cite{Smit2022Reverse}.\\

\begin{figure}
    \centering
    \includegraphics[scale=0.35]{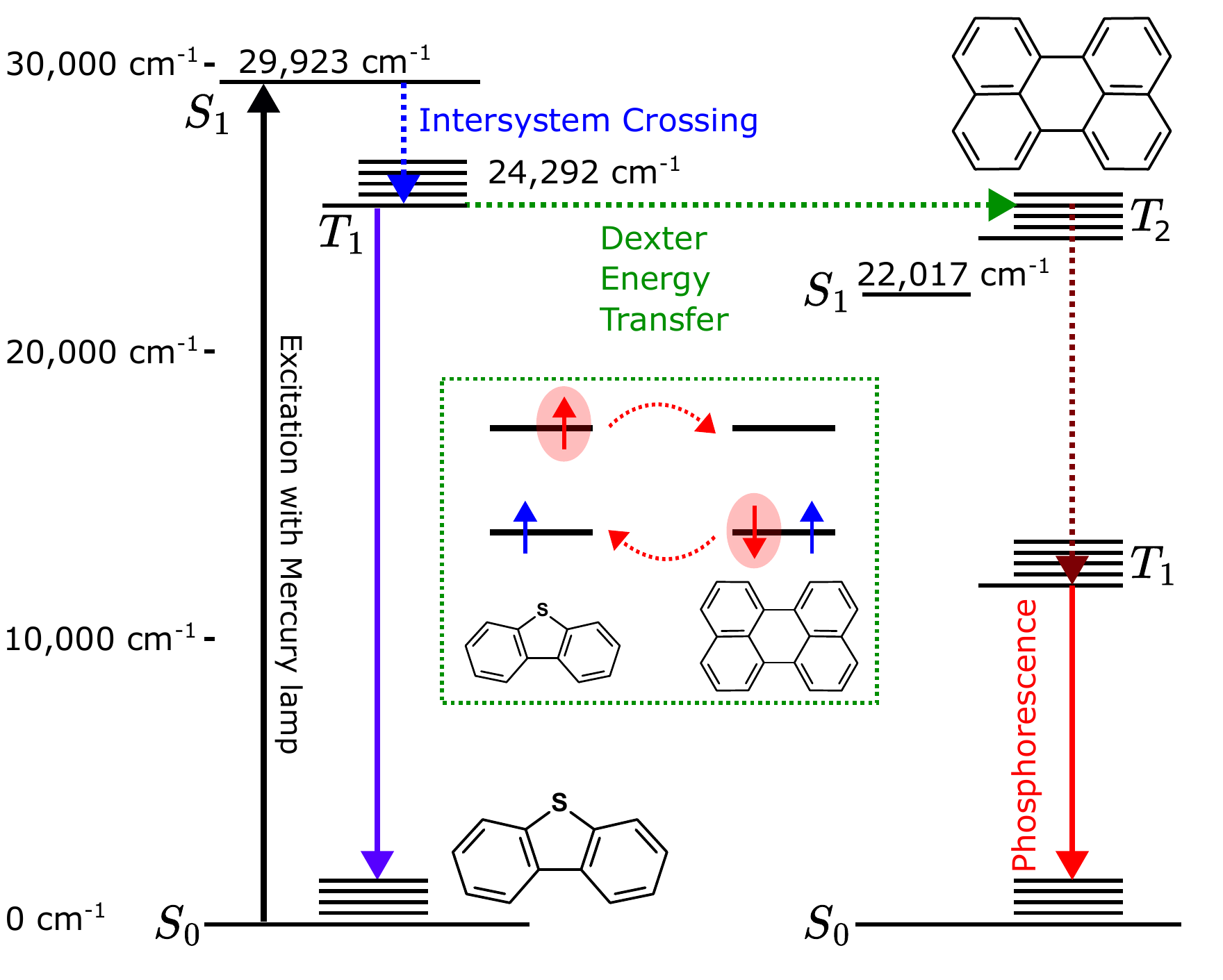}
    \caption{\textbf{Intermolecular enhancement scheme for perylene phosphorescence}. This Jablonski diagram depicts the energy levels of dibenzothiophene on the left side and perylene on the right side. To increase the rate of phosphorescence for perylene, the host material is excited by UV light from a Mercury lamp to $S_1$. Through intersystem crossing dibenzothiophene can then end up in the triplet state $T_1$. These non-bound triplets, i.e. excitons, diffuse by Dexter energy transfer until they are trapped in a lower energy state (X-trap), which can be for instance a perylene molecule. The Dexter energy transfer between dibenzothiophene and perylene is here shown by the hypothetical route via $T_2$ of perylene. In addition, $T_1$, can still be involved in the transfer process.}
    \label{Figure3}
\end{figure}

To overcome these limitations and increase triplet generation efficiency without compromising the phosphorescence signal from perylene, we explored whether the dibenzothiophene host could be exploited to produce triplet excitons capable of migrating to perylene molecules. This concept is illustrated in the Jablonski diagram of Figure \ref{Figure3}. Literature reports place the singlet and triplet states of dibenzothiophene at 29,923 cm$^{-1}$ (334.3 nm) and 24,292 cm$^{-1}$ (411.8 nm), respectively\cite{Bree1971Electronic}, both lying energetically above the singlet state of perylene (22,033 cm$^{-1}$ ; 454.0 nm) and thereby avoiding intermolecular intersystem crossing. The intersystem crossing rate of dibenzothiophene was found to be enhanced, possibly by a heavy-atom effect associated with its central sulfur atom\cite{Goldacker1979Electronic}, and as a result, dibenzothiophene exhibits relatively strong phosphorescence even at room temperature\cite{Zhao2019Boosting}.\\
\\
To populate triplet states in perylene impurities, energy transfer from the triplet state of dibenzothiophene can proceed via Dexter-type transfer\cite{Dexter1953Theory}, which relies on wavefunction overlap between donor and acceptor. However, this mechanism is typically efficient only when the energy gap between donor and acceptor triplet states is small\cite{Brown1982Triplet}. Previous studies demonstrated that a large gap reduces the probability of transfer, though increasing the concentration of trap molecules can partially compensate for this limitation\cite{Baessler1970Intermolecular}. Interestingly, quantum-chemical calculations suggest that perylene possesses a higher triplet state, $T_2$, located only $\sim$0.2 eV above its singlet state in vacuum\cite{Giri2018Model}. If this small energy gap is preserved upon incorporation into a dibenzothiophene matrix, the $T_2$ state could lie below the triplet state of dibenzothiophene and thus facilitate efficient Dexter transfer.\\
\\
When applying the excitation scheme of Figure \ref{Figure3} at low temperatures (<4 K), we observe an intense and long-lived phosphorescence from dibenzothiophene in the delayed emission signal. The recorded spectrum (Figure \ref{Figure4}a) reveals narrow emission lines arising from isolated defect sites within the dibenzothiophene crystal lattice, commonly referred to as X-traps\cite{Goldacker1979Electronic}. The vibrational structure of this spectrum closely matches previously reported data\cite{Bree1971Electronic}, although the observed linewidths ($\sim$8 $\pm$ 2 cm$^{-1}$) are somewhat broader than those reported for high-quality crystals ($\sim$1.5 cm$^{-1}$)\cite{Goldacker1979Electronic}, likely due to reduced crystal quality resulting from rapid quenching in liquid nitrogen.\\

\begin{figure}
    \centering
    \includegraphics[scale=0.37]{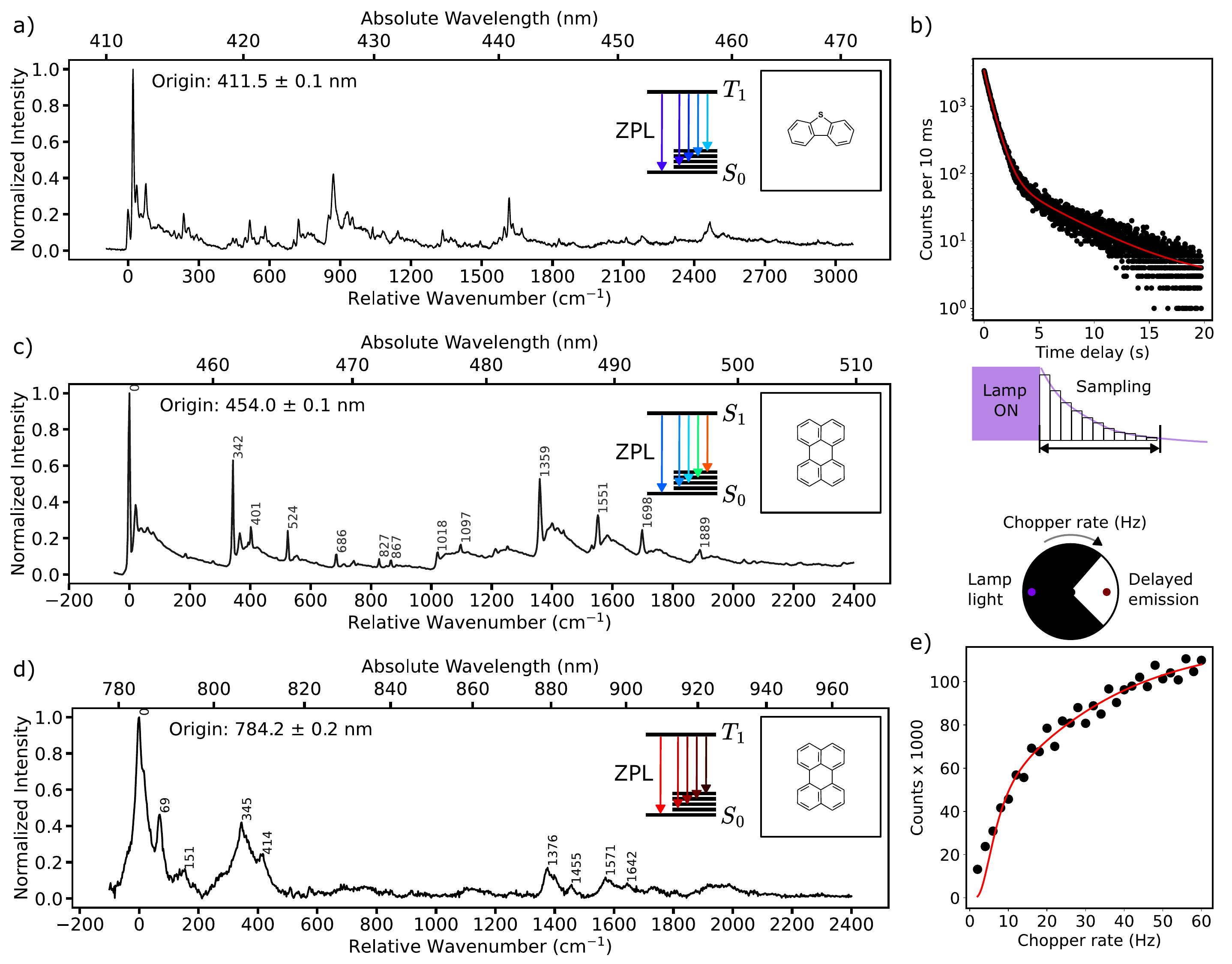}
    \caption{\textbf{Spectroscopic properties of perylene and dibenzothiophene}. The phosphorescence spectrum of dibenzothiophene is shown in panel (a). The intensity of the phosphorescence decreases exponentially over time, measured by sampling at 100 Hz, after switching off the lamp, as shown in panel (b). The fluorescence and phosphorescence spectrum of perylene are compared in panel (c) and (d). The phosphorescence decay is due to mix-up with other signals measured indirectly by varying the optical chopper rate and then fitted with equation 8 from the Supplementary. The post-processing steps required to obtain this spectrum are explained in Supplementary Figure S5.}
    \label{Figure4}
\end{figure}

Phosphorescence decay kinetics (Figure \ref{Figure4}b), measured with a single-photon counter after mechanically blocking the excitation light, were fitted to a triple-exponential function, yielding time constants of 0.24 $\pm$ 0.03 s, 0.68 $\pm$ 0.01 s, and 4.7 $\pm$ 0.3 s. These values are consistent with literature reports\cite{Goldacker1979Electronic}. Notably, as the temperature increases above 4 K, the dibenzothiophene phosphorescence signal diminishes rapidly. This behavior contrasts with undoped dibenzothiophene, where phosphorescence persists up to 77 K\cite{Goldacker1979Electronic}, and is likely due to transfer of the triplets to impurities.\\
\\
Surprisingly, the delayed emission spectra vary significantly between samples, depending on the section of the zone-refining tube from which the dibenzothiophene was sourced. This variation is likely due to presence of concentration gradients of impurities introduced during zone-refining. Among the impurities, we identified fluorene by its zero-phonon line at 424 nm and associated vibrational fingerprint (Supplementary Figure S3), along with two unidentified impurities producing red-shifted emission at 476 nm and 672 nm (Supplementary Figure S4). Notably, the latter result demonstrates that it is possible to observe emission from impurities bridging a substantial energy gap between the host's triplet and impurities' triplet states.\\
\\
Only material collected from the most purified section of the zone-refining tube yielded a phosphorescence spectrum consistent with that of perylene. This phosphorescence spectrum, integrated over 45 minutes (Figure \ref{Figure4}d), exhibits a clear zero-phonon line and a vibronic progression involving the 346 cm$^{-1}$ long-axis stretching mode of perylene, also observed in the fluorescence spectrum (Figure \ref{Figure4}c). The main zero-phonon line, corresponding to the $T_1\rightarrow S_0$ transition, is located at 12,756 cm$^{-1}$ (784.2 nm), calibrated against krypton lamp emission lines. Several sidebands shifted by approximately 69 cm$^{-1}$, namely 345 cm$^{-1}$ and 414 cm$^{-1}$, 1376 cm$^{-1}$, and 1455 cm$^{-1}$ are observed. Notably, some of these sidebands, absent in the fluorescence spectrum, likely originate from different spectroscopic sites identified in Figure \ref{Figure2}. For instance, the fluorescence site at 454.1 nm (Site 2) may correspond to phosphorescence at 788.4 nm, while the weak feature at 793.2 nm could relate to the red-shifted site at 455.5 nm (Site 3). We note that the linewidths of spectral features in the phosphorescence spectrum ($\sim$30 cm$^{-1}$) are considerably larger than for the fluorescence spectrum ($\sim$10 cm$^{-1}$). This could be the result of multiple factors: namely the high concentration of perylene used for the phosphorescence experiment, a narrowband excitation source was used to obtain the fluorescence spectrum and the phosphorescence emission was collected from a much larger area. All of them can explain the inhomogeneity observed in the phosphorescence spectrum. Most importantly, inhomogeneities are filtered out in the fluorescence spectrum by the narrowband excitation, whereas the energy transfer to perylene molecules likely occurs over a broad range of energies, covering all sites.\\
\\
By varying the chopper wheel frequency, we controlled the delay between the excitation pulse and detection window, set at one-quarter of the chopper cycle. The phosphorescence decay was analyzed by integrating the signal over different sections of the decay curve as a function of the chopper rate. The resulting data were fitted to a model described in Supplementary Equation 8. The best fit was obtained with two phosphorescence lifetimes, 4.9 $\pm$ 1.2 ms and 27 $\pm$ 4 ms, slightly shorter than those previously observed in single-molecule experiments (8.5 $\pm$ 0.4 ms and 64 $\pm$ 12 ms)\cite{Smit2022Reverse}. This difference likely originates from ensemble averaging effects and contributions from multiple spectroscopic sites with distinct lifetimes, whereas the corresponding single-molecule measurements have only been measured for a single site and exhibit significant molecule-to-molecule variability.\\
\\
Finally, we establish an upper bound for the triplet energy transfer efficiency from dibenzothiophene to perylene. Under identical spectrometer settings and a setup corrected for chromatic aberration, dibenzothiophene phosphorescence at low temperature yielded approximately 7.7 million counts per second, compared to only 1000 counts per second for perylene. Accounting for differences in phosphorescence quantum yields — estimated from triplet lifetimes and a radiative lifetime of 30 s to be 0.8–1.6\% for the $T_x$ and $T_y$ states and 16\% for the $T_z$ state of dibenzothiophene, consistent with room-temperature values of 2–3\%\cite{Fang2018White} — we estimate the phosphorescence quantum yield of perylene to be 0.03\% for the $T_x$ and $T_y$ states and 0.2\% for the $T_z$ state. Assuming comparable detection efficiencies for blue and near-infrared emission, we derive an upper limit of approximately 1\% for the triplet energy transfer efficiency. In practice, this value may be somewhat lower due to the higher detection efficiency for near-infrared photons. Nevertheless, considering the substantial energy gap between the triplet states of dibenzothiophene and perylene, this relatively high efficiency suggests the possible involvement of an intermediate triplet state, such as $T_2$, in the energy transfer dynamics.\\
\\
We note that the significant energy gap between $T_2$ and $T_1$ could promote triplet-triplet fluorescence over nonradiative decay. To exclude that the observed emission at 784.2 nm is a signature of delayed triplet fluorescence, we argue that the total energy of the exciton of dibenzothiophene is 24,292 cm$^{-1}$ and minus the observed zero phonon line at 12,756 cm$^{-1}$ this leaves an energy of 11,536 cm$^{-1}$, equal to 867 nm or higher if part of the energy is transformed into heat. This wavelength is far off from the expected wavelength for the $S_0$-$T_1$ energy gap. Moreover, as we detect a millisecond-range phosphorescence decay, recorded at delays of several milliseconds, we would effectively filter out the likely short-lived $T_2$-$T_1$ fluorescence. If the long-lived triplet of dibenziothiophene would supply excitons during its entire lifetime - on the order of seconds - the delayed luminescence would have a decay of longer than milliseconds. This argument also agrees with our assumption that the transfer of triplet excitons happens instantaneously compared to the triplet lifetime of perylene (used to obtain equation 8 in the Supplementary). \\
\\
With the triplet's identified wavelength of 784.2 nm, an inhomogeneous broadening of 30 cm$^{-1}$, and precision of about 3 cm$^{-1}$, we have substantially narrowed the spectral region for locating a dark resonance to the triplet state. Further improvements in spectral precision may be achieved by enhancing crystal quality through sublimation growth, which could reduce the inhomogeneous broadening of spectroscopic sites below 1 cm$^{-1}$\cite{brouwer_single-molecule_1999}. As a logical next step, we propose a targeted experiment to directly identify the dark resonance of a single molecule by monitoring its fluorescence intensity while scanning a near-infrared laser across the refined wavelength region. 

\section*{Conclusion}

In this work we have demonstrated that the detection bottleneck for the ultra-weak phosphorescence of guest molecules can be solved by using the host material as a source for triplet excitons. Moreover, we found that the efficiency of triplet energy transfer strongly depends on crystal purity, in order to make the intentional impurities, i.e. guest molecules, the dominant acceptor for triplet excitons. Our used method holds promise for application in other host-guest systems, in particular for hosts that can strongly phosphoresce themselves. A likely candidate is terrylene in BTBT\cite{Smit2024Probing}, having a similar structure to dibenzothiophene and displaying phosphorescence by UV excitation\cite{Deperasinska2021Spectra}. With our experiment we have laid a basis for future experiments that continue with the perylene in dibenzothiophene system for finding the weak resonance to the triplet manifold, to aid all-optical spin manipulation in single molecules. 

\section*{DATA AVAILABILITY}
The data that support the findings of this study are available from the corresponding author upon reasonable request.

\section*{ACKNOWLEDGMENTS}
We are grateful for the funding provided by NWO (Spinoza prize 2017) that made this work possible.  

\clearpage
\renewcommand{\thesection}{S\arabic{section}}
\renewcommand{\thefigure}{S\arabic{figure}}
\renewcommand{\thetable}{S\arabic{table}}
\renewcommand{\theequation}{S\arabic{equation}}
\setcounter{section}{0}
\setcounter{figure}{0}
\setcounter{table}{0}
\setcounter{equation}{0}

\section*{Supplementary Material}
\addcontentsline{toc}{section}{Supplementary Material}

\sisetup{exponent-product=\cdot}

\renewcommand{\thefigure}{S\arabic{figure}}

\title{Supplementary information for: Towards all-optical spin manipulation in single molecules: a refined region for locating a dark resonance}

\date{June 2025}

\author{Robert Smit}\email{rsmit@physics.leidenuniv.nl}
\affiliation{Huygens-Kamerlingh Onnes Laboratory, LION, Postbus 9504, 2300 RA Leiden, The Netherlands}
\author{Boleslaw Kozankiewicz}
\affiliation{Institute of Physics, Polish Academy of Sciences Al. Lotnikow 32/46, 02-668 Warsaw, Poland}
\author{Michel Orrit}
\affiliation{Huygens-Kamerlingh Onnes Laboratory, LION, Postbus 9504, 2300 RA Leiden, The Netherlands}

\maketitle

\section{Time-gated detection of phosphorescence}\label{SM_A}

The measurement setup for the detection of phosphorescence – the phosphoroscope – is shown in Figure \ref{Figure1}. The excitation light is generated by a mercury lamp and is filtered to transmit only the UV part below 400 nm (see spectrum on the top of Figure \ref{Figure1}). The chopper wheel separates the excitation and detection windows by a fourth of a cycle. The optical paths for the excitation and emission light are separated such that they pass opposite sides of the optical chopper. \\

\begin{figure}
    \centering
    \includegraphics[scale=0.27]{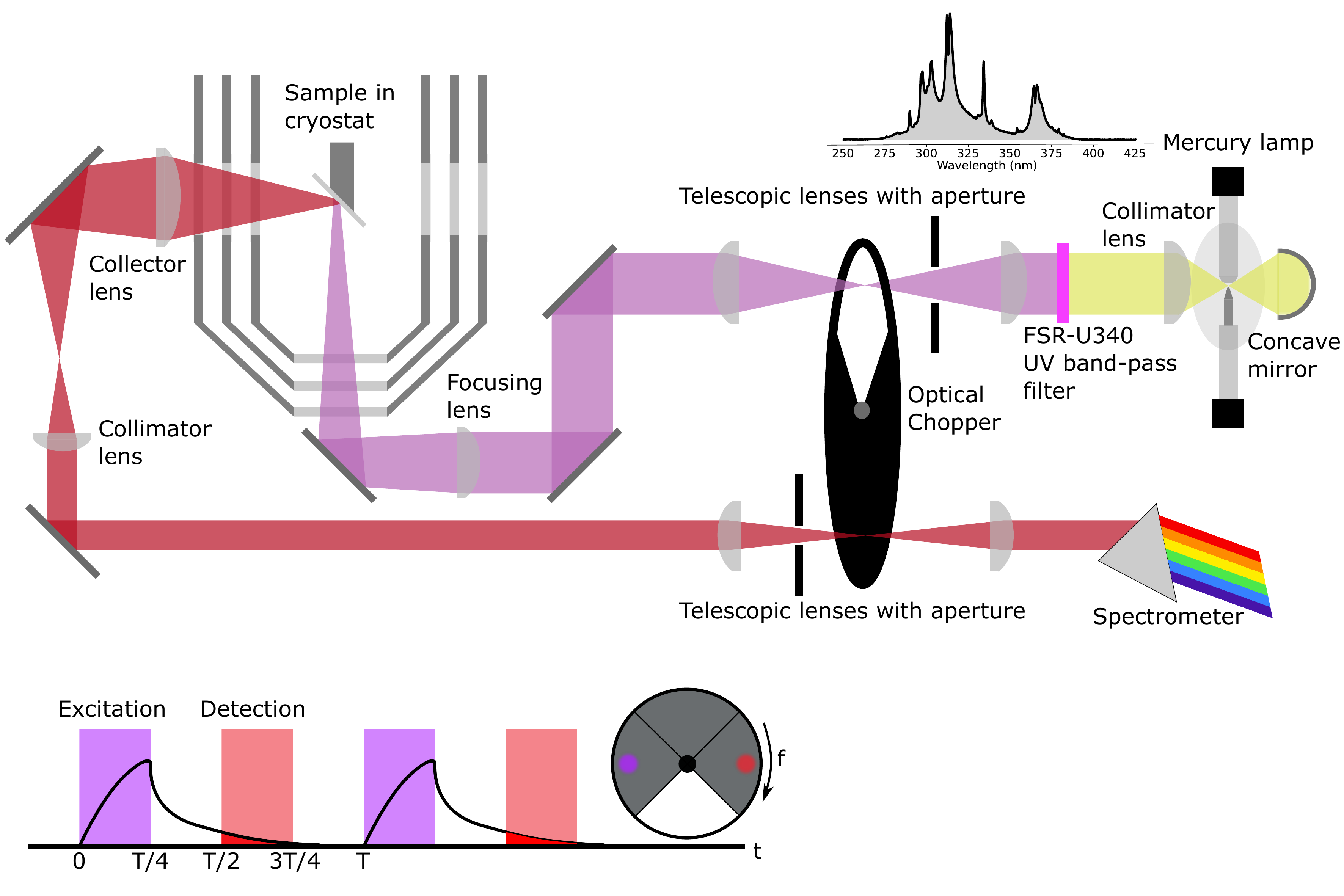}
    \caption{Schematic of the setup used for detecting delayed emission signals. Not shown is the extensive shielding that was present to eliminate stray light entering into the path of the detector. A schematic on the bottom shows the measurement sequence of the chopper.}
    \label{Figure1}
\end{figure}
\newpage
The accumulated phosphorescence signal $S$, as a function of chopper frequency $f$, is derived from rate equations, solved for the steady state. For simplification we start with a single rate of decay from the triplet state and we assume that relaxation from the singlet excited state of dibenzothiophene to the triplet state of perylene (i.e. intersystem crossing followed by Dexter energy transfer) is instantaneous compared to the triplet lifetime. Following these assumptions, there are two rate parameters of concern: $k_{ST}$ for the rate of excitation to the triplet and $k_{TS}$ for the rate of decay from the triplet. As the excitation is on, the population of the triplet $p$ builds up as: 

\begin{equation}
\dot{p} = k_{ST} - k_{TS}p. 
\end{equation}

In steady state – equivalent to a continuously rotating chopper – the differential equation approaches the value $p_s$ = $k_{ST}$/$k_{TS}$. A general solution for equation 1 is therefore: 

\begin{equation}
p(t) = p_s + Ae^{-k_{TS}t}. 
\end{equation}

At $t=0$ equation 2 amounts to $p(0)=p_s+A= p_0$. Replacing A by the expression for $p_0$ we end up with the expression for p(t): 

\begin{equation}
p(t)= p_0 e^{-k_{ST} t}+ p_s (1-e^{-k_{ST} t}).   
\end{equation}

At $t=T/4$ we have thus accumulated a triplet population given by $p(T/4)$. At this point in time, the acquired population will decay naturally as: 

\begin{equation}
p_d(t)=p(T/4) e^{-k_{TS} (t-\frac{T}{4})}.
\end{equation}

The detection of the phosphorescence signal occurs from $t = T/2$ up to $t = 3T/4$ (Figure S1) and thus the accumulated signal $S$ at the detector is given by an integral over $p_d(t)$: 

\begin{equation}
S= \int_{T/2}^{3T/4} p_d(t)dt = \frac{p(T/4)}{k_{TS}}e^{\frac{k_{TS}T}{4}} (e^{-k_{TS}T/2}-e^{-k_{TS} 3T/4}). 
\end{equation}

Due to periodicity of the chopper, thus $p(0)=p(T)$, we can solve for $p_0$ and find an expression for it: 

\begin{equation}
p_0 = p_s \frac{e^{k_{TS} T/4}-1}{e^{k_{TS} T}-1}.
\end{equation}

Finally, the signal per chopper period T amounts to: 

\begin{equation}
\frac{S}{T}=  \frac{k_{ST}}{(k_{TS}^2 T)}  \frac{1-e^{-k_{TS} T/4}}{1-e^{-k_{TS} T }} (e^{-k_{TS}(\frac{T}{2}-\frac{T}{4})} - e^{-k_{TS} (\frac{3T}{4}-\frac{T}{4})}).    
\end{equation}

For fitting the data in Figure 4e in the main text, we take the sum of equation 7 for two different values of $k_{TS}$. With arbitrary scaling factors $C_i$, the equation then reduces to: 

\begin{equation}
S_1 + S_2 = \sum_{i=i}^{2} C_i*\frac{1-e^{-1/4f\tau_{i}}}{1-e^{-1/f\tau_{i}}}(e^{-\frac{1}{\tau_i}(\frac{1}{2f} - \frac{1}{4f})} - e^{-\frac{1}{\tau_i}(\frac{3}{4f} - \frac{1}{4f})})
\end{equation}

\section{Delayed emission spectra}

\begin{figure}
    \centering
    \includegraphics[scale=0.4]{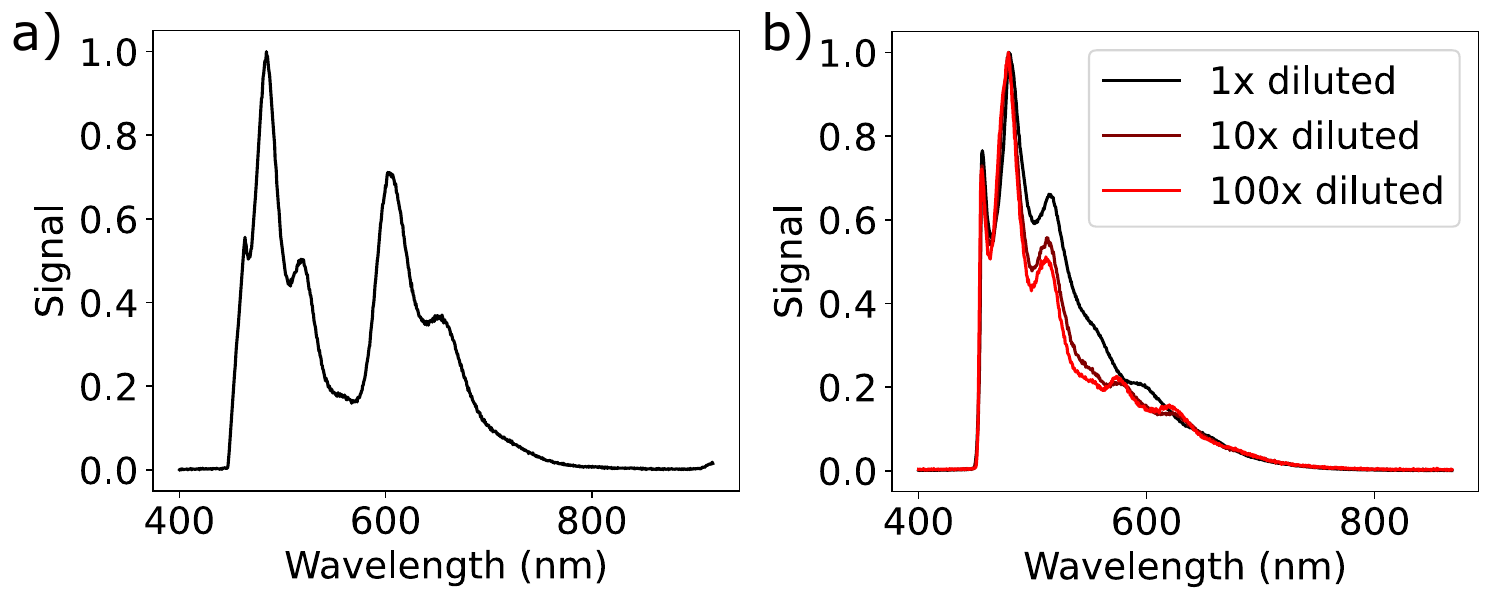}
    \caption{Panel (a) shows the fluorescence emission spectrum of perylene in dibenzothiophene for a very high concentration (in the parts per thousand range). There are strong emission features around 600-700 nm, assigned to Pr excimer emission. Panel (b) shows emission spectra for a lower concentration than in (a) and is compared to increasingly lower concentrations. The large peaks are not visible anymore, but further decrease of intensity is noticed on the red side of the perylene spectrum. All spectra were taken at room temperature with a 405 nm excitation laser and filtered with a 460 nm Long-pass filter.}
    \label{Figure2}
\end{figure}

To obtain a detectable phosphorescence signal from perylene, the concentration of perylene must exceed those of other traps, in order to increase the likelihood of trapping the triplet excitons at perylene defect sites. However, the concentration cannot be increased without cost. Figure \ref{Figure2}a shows that for a very high concentration - the crystals turned yellow - new features appear in the red part of the emission spectrum. This red emission is typically associated with excimer emission due to perylene dimers\cite{ferguson_absorption_1966}. For increasingly lower concentrations, as shown in Figure \ref{Figure2}b, the red part of the spectrum becomes less pronounced.  \\

\begin{figure}
    \centering
    \includegraphics[scale=0.36]{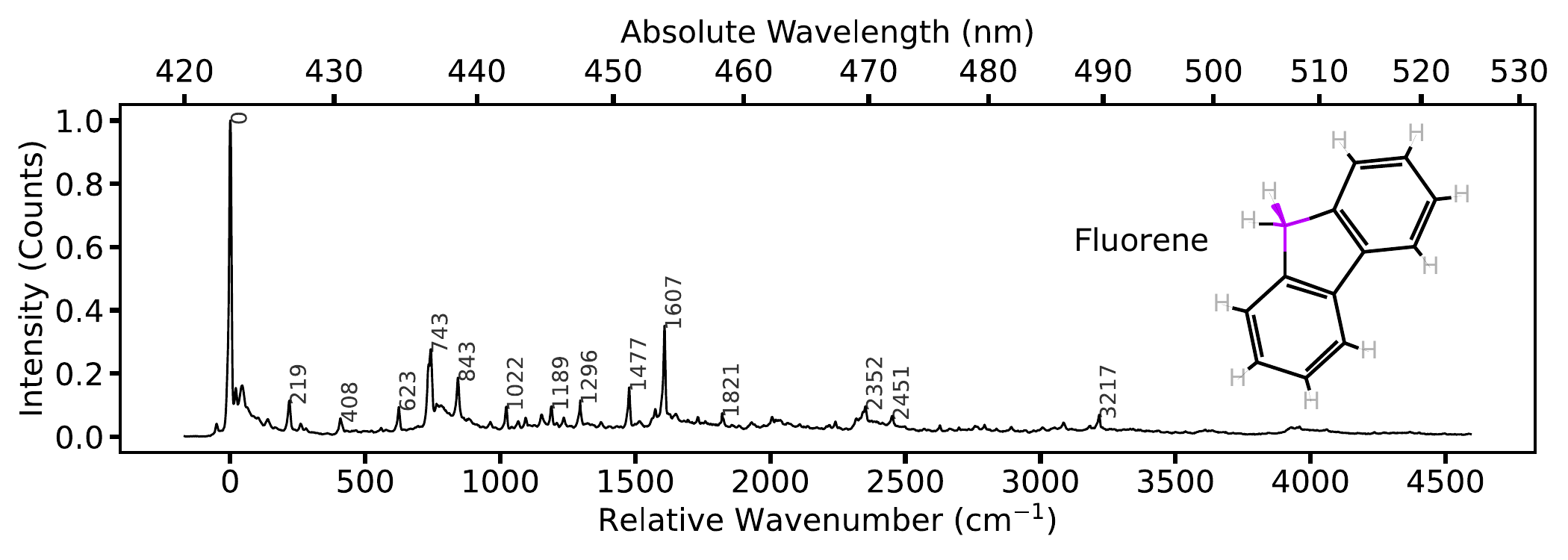}
    \caption{Delayed emission signal observed in one of the experiments with less pure material from the zone-refining tube. The identified impurity, fluorene, is shown with its structure in the inset.}
    \label{Figure3}
\end{figure}

The measurements of different samples revealed different impurities that dominated in the delayed emission signal. In the case of Figure \ref{Figure3}, the recorded emission spectrum can be assigned to the presence of fluorene\cite{bree_study_1969}. Other unidentified emission spectra are shown in Figure \ref{Figure4}. \\

\begin{figure}
    \centering
    \includegraphics[scale=0.36]{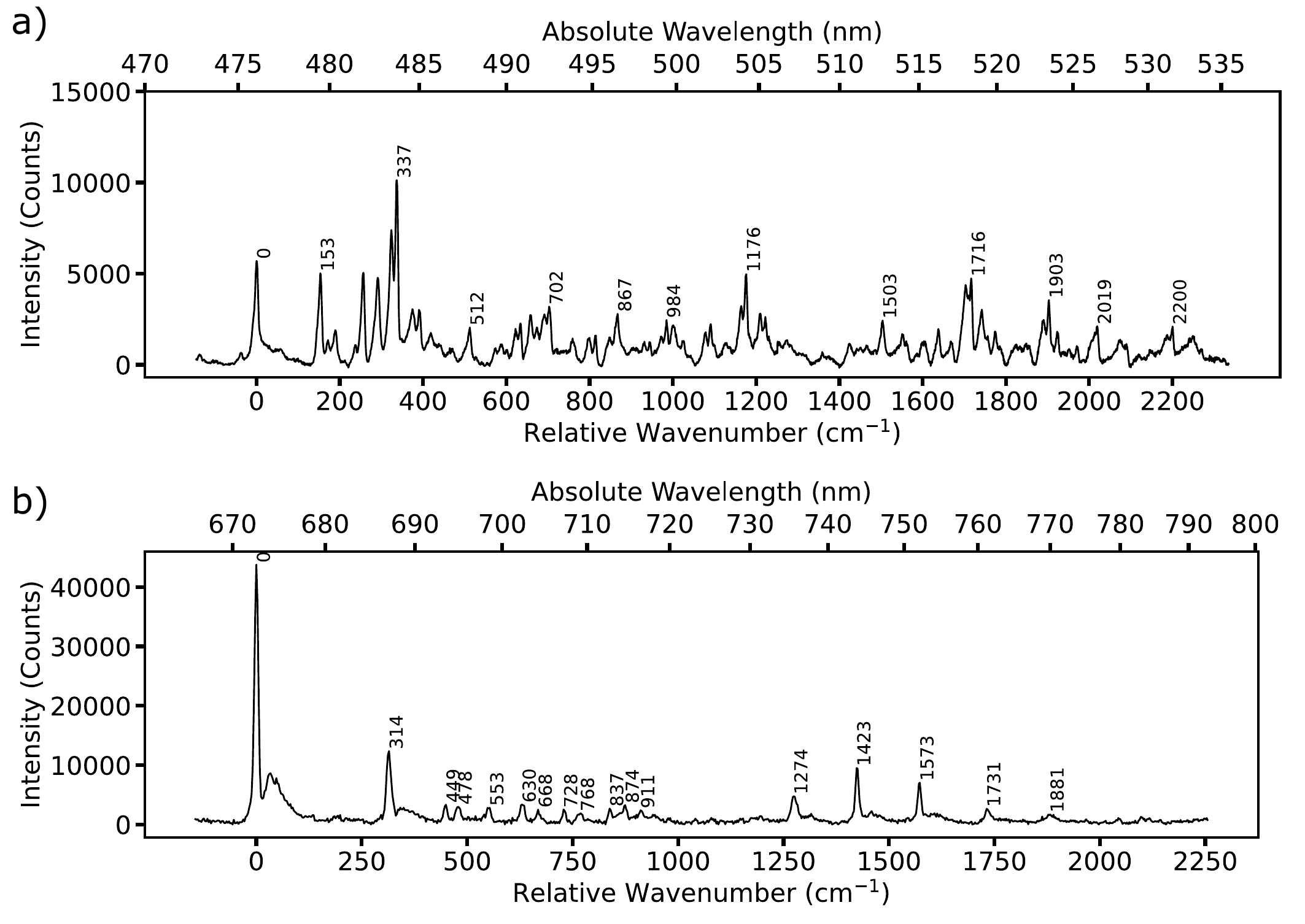}
    \caption{Another set of delayed emission spectra recorded from different samples. The particular sources of the emission are not identified. Apart from perylene, the spectrum in (b) is the most red-shifted emission recorded from the samples. }
    \label{Figure4}
\end{figure}

\begin{figure}
    \centering
    \includegraphics[scale=0.6]{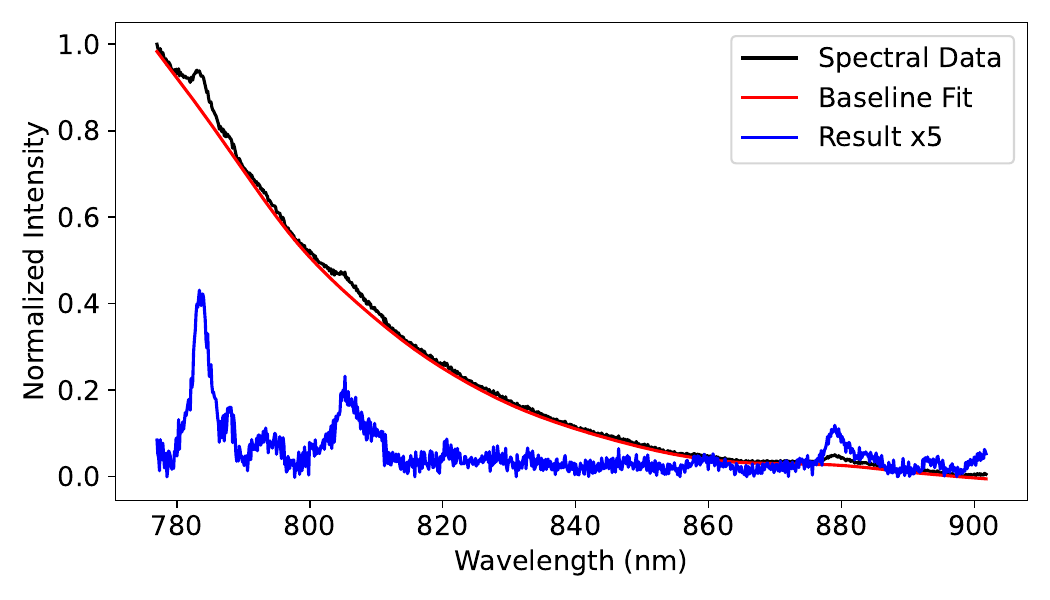}
    \caption{Typical delayed emission spectrum in the near-infrared region at $f$ = 18 Hz. The spectrum of perylene (blue curve), such as for the spectrum in Figure 4d in the main text, has been filtered by removing background luminescence from the recorded spectrum (black curve) by subtracting a fitted baseline (red curve).}
    \label{Figure5}
\end{figure}

The delayed emission was typically accompanied by a strong background signal. This is for instance the case for the near-infrared region where the phosphorescence of perylene was detected, as shown in Figure \ref{Figure5}. To remove this background luminescence, the spectrum was fitted to a smooth baseline function. This baseline was then subtracted from the data to obtain the structured phosphorescence spectrum. \\

\section*{REFERENCES}
\bibliography{refs}

\end{document}